\documentclass[twocolumn,showpacs,aps,prl,superscriptaddress]{revtex4}

\usepackage{graphicx}
\usepackage{dcolumn}
\usepackage{amsmath}
\usepackage{epsfig}
\usepackage{units}

\RequirePackage{xspace}

\usepackage{relsize}
\def\babar{\mbox{\slshape B\kern-0.1em{\smaller A}\kern-0.1em
    B\kern-0.1em{\smaller A\kern-0.2em R}}}

\def\etal   {{\it et al.}}

\def\epem       {\ensuremath{e^+e^-}\xspace}

\def\piz   {\ensuremath{\pi^0}\xspace}

\def\pip   {\ensuremath{\pi^+}\xspace}
\def\pim   {\ensuremath{\pi^-}\xspace}

\def\Kbar  {\kern 0.2em\overline{\kern -0.2em K}{}\xspace}

\def\Kz    {\ensuremath{K^0}\xspace}
\def\Kzb   {\ensuremath{\Kbar^0}\xspace}
\def\KzKzb {\ensuremath{\Kz \kern -0.16em \Kzb}\xspace}
\def\Kp    {\ensuremath{K^+}\xspace}
\def\Km    {\ensuremath{K^-}\xspace}

\def\KpKm  {\ensuremath{\Kp \kern -0.16em \Km}\xspace}
\def\KS    {\ensuremath{K^0_{\scriptscriptstyle S}}\xspace} 
 
\def\Kstarz  {\ensuremath{K^{*0}}\xspace}

\def\Kstar   {\ensuremath{K^*}\xspace}

\def\Dbar    {\kern 0.2em\overline{\kern -0.2em D}{}\xspace}

\def\Dz      {\ensuremath{D^0}\xspace}
\def\Dzb     {\ensuremath{\Dbar^0}\xspace}
\def\DzDzb   {\ensuremath{\Dz {\kern -0.16em \Dzb}}\xspace}
\def\Dp      {\ensuremath{D^+}\xspace}
\def\Dm      {\ensuremath{D^-}\xspace}

\def\DpDm    {\ensuremath{\Dp {\kern -0.16em \Dm}}\xspace}

\def\B       {\ensuremath{B}\xspace}
\def\Bbar    {\kern 0.18em\overline{\kern -0.18em B}{}\xspace}

\def\BB      {\ensuremath{B\Bbar}\xspace} 
\def\Bz      {\ensuremath{B^0}\xspace}
\def\Bzb     {\ensuremath{\Bbar^0}\xspace}
\def\BzBzb   {\ensuremath{\Bz {\kern -0.16em \Bzb}}\xspace}
\def\Bu      {\ensuremath{B^+}\xspace}
\def\Bub     {\ensuremath{B^-}\xspace}

\def\BpBm    {\ensuremath{\Bu {\kern -0.16em \Bub}}\xspace}

\mathchardef\Upsilon="7107
\def\Y#1S{\ensuremath{\Upsilon{(#1S)}}\xspace}

\mathchardef\Deltares="7101
\mathchardef\Xi="7104
\mathchardef\Lambda="7103
\mathchardef\Sigma="7106
\mathchardef\Omega="710A

\def\Deltabar{\kern 0.25em\overline{\kern -0.25em \Deltares}{}\xspace}
\def\Lbar{\kern 0.2em\overline{\kern -0.2em\Lambda\kern 0.05em}\kern-0.05em{}\xspace}
\def\Sigbar{\kern 0.2em\overline{\kern -0.2em \Sigma}{}\xspace}
\def\Xibar{\kern 0.2em\overline{\kern -0.2em \Xi}{}\xspace}
\def\Obar{\kern 0.2em\overline{\kern -0.2em \Omega}{}\xspace}
\def\Nbar{\kern 0.2em\overline{\kern -0.2em N}{}\xspace}
\def\Xb{\kern 0.2em\overline{\kern -0.2em X}{}\xspace}

\def\mes        {\mbox{$m_{\rm ES}$}\xspace}

\def\DeltaE     {\mbox{$\Delta E$}\xspace}

\newcommand{\tev}{\ensuremath{\mathrm{\,Te\kern -0.1em V}}\xspace}
\newcommand{\gev}{\ensuremath{\mathrm{\,Ge\kern -0.1em V}}\xspace}
\newcommand{\mev}{\ensuremath{\mathrm{\,Me\kern -0.1em V}}\xspace}
\newcommand{\kev}{\ensuremath{\mathrm{\,ke\kern -0.1em V}}\xspace}
\newcommand{\ev}{\ensuremath{\mathrm{\,e\kern -0.1em V}}\xspace}
\newcommand{\gevc}{\ensuremath{{\mathrm{\,Ge\kern -0.1em V\!/}c}}\xspace}
\newcommand{\mevc}{\ensuremath{{\mathrm{\,Me\kern -0.1em V\!/}c}}\xspace}
\newcommand{\gevcc}{\ensuremath{{\mathrm{\,Ge\kern -0.1em V\!/}c^2}}\xspace}
\newcommand{\mevcc}{\ensuremath{{\mathrm{\,Me\kern -0.1em V\!/}c^2}}\xspace}

\newcommand{\zrec}{\ensuremath{z_{\CP}}}
\newcommand{\ztag}{\ensuremath{z_\mathrm{tag}}}

\def\mus  {\ensuremath{\rm \,\mus}\xspace}

\def\mus        {\ensuremath{\,\mu{\rm s}}\xspace}    

\def\ra                 {\ensuremath{\rightarrow}\xspace}
\def\to                 {\ensuremath{\rightarrow}\xspace}

\def\pep2{PEP-II}

\def\gsim{{~\raise.15em\hbox{$>$}\kern-.85em
          \lower.35em\hbox{$\sim$}~}\xspace}
\def\lsim{{~\raise.15em\hbox{$<$}\kern-.85em
          \lower.35em\hbox{$\sim$}~}\xspace}

\def\CP                {\ensuremath{C\!P}\xspace}

\def\deltaz{\ensuremath{{\rm \Delta}z}\xspace}
\def\deltat{\ensuremath{{\rm \Delta}t}\xspace}
\def\deltamd{\ensuremath{{\rm \Delta}m_d}\xspace}

\xspace

\newcommand{\jprlBase}       {Phys.\ Rev.\ Lett.\xspace}
\newcommand{\jprBase}        {Phys.\ Rev.\xspace}

\newcommand{\nimBaseC}       {Nucl.\ Instr.\ and Methods\xspace}

\newcommand{\nima}      [1]  {\nimBaseC~A~{\bf #1}}

\newcommand{\jprl}      [1]  {\jprlBase\ {\bf #1}}
\newcommand{\jprd}      [1]  {\jprBase\ D~{\bf #1}}

\newcommand{\progtp}    [1]  {{Prog.\ Th.\ Phys.\ {\bf #1}}}

\def\jetset74   {\mbox{\tt Jetset \hspace{-0.5em}7.\hspace{-0.2em}4}\xspace}

\def\microns {\ensuremath{\mu{\rm m}}}
\def\Btag {\ensuremath{\B_{\rm tag}}}

\def\Bztokspiz {\ensuremath{\Bz \to \KS\piz}}

\def\Bztokstargamma {\ensuremath{\Bz \to K^{*0}\gamma}}

\def\ckstargamma {\ensuremath{C_{\Kstar\gamma}}}
\def\skstargamma {\ensuremath{S_{\Kstar\gamma}}}
\def\cf {\ensuremath{C_f}}
\def\sf {\ensuremath{S_f}}
\def\fish    {\ensuremath{\cal F}}

\def\Btag {\ensuremath{B_{\rm tag}}}

\def\TheDecay {\ensuremath{\Bztokstargamma  (\Kstarz\ra\KS\piz)}}

\def\figurebox#1#2#3{%
    \def\arg{#3}%
    \ifx\arg\empty
    {\hfill\vbox{\hsize#2\hrule\hbox to #2{\vrule\hfill\vbox to #1{\hsize#2\vfill}\vrule}\hrule}\hfill}%
    \else
    {\hfill\epsfbox{#3}\hfill}%
    \fi}

\long\def\inst#1{\par\nobreak\kern 4pt\nobreak
    {\it #1}\par\vskip 10pt plus 3pt minus 3pt}

\begin{document}


\title{
{ \Large \bf \boldmath Measurement of Time-dependent \CP-violating Asymmetries in
\TheDecay\ Decays }
}

%
\author{B.~Aubert}
\author{R.~Barate}
\author{D.~Boutigny}
\author{F.~Couderc}
\author{J.-M.~Gaillard}
\author{A.~Hicheur}
\author{Y.~Karyotakis}
\author{J.~P.~Lees}
\author{V.~Tisserand}
\author{A.~Zghiche}
\affiliation{Laboratoire de Physique des Particules, F-74941 Annecy-le-Vieux, France }
% ...
\collaboration{The \babar\ Collaboration}
\noaffiliation


\date{\today}

\begin{abstract}
  We present a measurement of the time-dependent \CP-violating
  asymmetries in \TheDecay\ decays based on 124 million $\Y4S\to\BB$
  decays collected with the \babar\ detector at the PEP-II
  asymmetric-energy $B$ Factory at the Stanford Linear Accelerator
  Center. In a sample containing $105\pm 14$ signal decays, we measure
  $\skstargamma = 0.25 \pm 0.63 \pm 0.14$ and $\ckstargamma = -0.57
  \pm 0.32 \pm 0.09$, where the first error is statistical and the
  second systematic.
\end{abstract}


\maketitle
The recent data\cite{Sin2betaObs} from the $B$ factory experiments
have provided strong evidence that the quark mixing mechanism in the
Standard Model (SM), encapsulated in the Cabibbo-Kobayashi-Maskawa
(CKM) matrix\cite{CKM}, is the dominant source of \CP\ violation in
the quark sector.  Nonetheless, decays which originate from radiative
loop processes, such as $b\to s \gamma$, may exhibit significant
deviations from the SM due to new physics contributions. In this
letter we report the first measurement of time-dependent \CP-violating
(CPV) asymmetries in a $b\to s \gamma$ process through the exclusive
decay $B^0\to K^{*0}\gamma$, where
$\Kstar\to\KS\piz$\cite{ref:cc}. D. Atwood, M. Gronau and A. Soni were
the first to point out that such a measurement probes the polarization
of the photon \cite{soni}, which is dominantly left-handed
(right-handed) for $b\to s\gamma$ ($\bar b\to \bar s\gamma$) in the
SM, but is mixed in various new physics scenarios.  The exclusive
decays $B^0\to (\KS\piz)\gamma_R$ and $\bar{B}^0\to (\KS\piz)\gamma_L$
are orthogonal transitions and are the dominant decays in the SM.
Therefore the CPV asymmetry due to interference between decays with or
without mixing is expected to be very small, $\approx 2 (m_s/m_b) \sin
2\beta$ ($\beta\equiv\arg (-V_{cd}V^*_{cb}/V_{td}V^*_{tb})$).  Any
significant deviation would indicate phenomena beyond the SM.

The \Bztokstargamma\ decays have been previously explored by the
CLEO\cite{CLEOBRPRL}, \babar\cite{BabarBRPRL}, and Belle
collaborations \cite{BelleBR}, who reported measurements of branching
fractions and the direct \CP\ and isospin asymmetries.  The
measurements reported in this letter are based on 124 million
$\Y4S\to\BB$ decays collected in 1999-2003 at the PEP-II $e^+e^-$
collider at the Stanford Linear Accelerator Center with the \babar\
detector, which is fully described in Ref.~\cite{ref:babar}.  For the
extraction of the time dependence of \TheDecay\ decays, we adopt an
analysis approach that closely follows our recently published
measurement of CPV asymmetries in the decay $\Bz\to\KS\piz$
\cite{BaBarKsPi0}. There we established a technique of vertex
reconstruction for $B$ decay modes to final states containing a
$\KS\ra\pip\pim$ decay and other neutral particles, but no primary
charged particles at the $B$ decay vertex.

We search for \TheDecay\ decays in hadronic events, which are selected
based on charged particle multiplicity and event topology.  We
reconstruct $\KS\to\pip\pim$ candidates from pairs of oppositely
charged tracks, detected in the silicon vertex detector (SVT) and/or
the central drift chamber (DCH). We require that these tracks
originate from a vertex which is more than $0.3$~cm from the primary
vertex and that the resulting candidates have a $\pip\pim$ invariant
mass between $487$ and $508$\mevcc.  We form $\piz\to\gamma\gamma$
candidates from pairs of photon candidates in \babar's electromagnetic
calorimeter (EMC) which are not associated with any charged tracks,
carry a minimum energy of 30\mev, and possess the expected lateral
shower shape. We require that the $\gamma\gamma$ combination has an
energy greater than $200$\mev and an invariant mass between $115$ and
$155$\mevcc.  We reconstruct candidate $\Kstar\to \KS\piz$ decays from
$\KS\piz$ combinations with invariant mass in the range $0.8
<M(\KS\piz)<1.0$\gevcc.  For photons originating from the $B$ decay,
we select clusters in the EMC which are isolated by 25~cm from all
other energy deposits and are inconsistent with $\piz\to \gamma\gamma$
or $\eta\to \gamma\gamma$ decays.

We identify \Bztokstargamma\ decays in $\Kstar\gamma$ combinations
using two nearly independent kinematic variables: the
energy-substituted mass $\mes=\sqrt{(s/2+{\bf p}_i\cdot{\bf
p}_B)^2/E_i^2-p^2_B}$ and the energy difference
$\DeltaE=E^*_B-\sqrt{s}/2$.  Here $(E_i,{\bf p}_i)$ and $(E_B,{\bf
p}_B)$ are the four-vectors of the initial $e^+e^-$ system and the $B$
candidate, respectively, $\sqrt{s}$ is the center-of-mass energy, and
the asterisk denotes th center-of-mass (CMS) frame.  For signal decays, the
\mes\ distribution peaks near the $B$ mass with a resolution of
$\approx3.5\mevcc$ and \DeltaE\ peaks near 0\mev with a resolution of
$\approx50$\mev. Both \mes\ and \DeltaE\ exhibit a low-side tail from
energy leakage in the EMC.  For the study of CPV asymmetries, we
consider candidates within $5.2<\mes<5.3\gevcc$ and
$|\DeltaE|<300\mev$, which includes the signal as well as a
large ``sideband'' region for background estimation.  When more than
one candidate is found in an event, we select the combination with the
$\piz$ mass closest to the nominal $\piz$ value, and if
ambiguity persists we select the combination with the $\KS$
mass closest to the nominal \KS\ value.

The sample of candidate events selected by the above requirements
contains significant background contributions from continuum $e^+e^-
\to q\bar{q}$ $(q=\{u,d,s,c\})$, as well as random combinations from
generic \BB\ decays.  We suppress both of these backgrounds by taking
advantage of the expected angular distribution of the decay products
of these processes. Angular momentum conservation restricts the
\Kstar\ meson in the \Bztokstargamma\ decay to transversely polarized
states, which leads to an angular distribution of $\sin^2\theta_H$ for
the decay products, where $\theta_H$ is the angle between the \KS\ and
the $B$ meson directions in the \Kstar\ rest frame. Monte Carlo studies
show that the background candidates peak near $\cos\theta_H =-1$. We
require $\cos\theta_H>-0.6$, resulting in rejection of
$68\%$ of $\BB$ and $48\%$ of continuum background candidates, while
retaining $91\%$ of the signal.

We exploit topological variables to further suppress the continuum
backgrounds, which in the CMS frame tend to retain the jet-like
features of the $q\bar{q}$ fragmentation process, as opposed to
spherical \BB\ decays.  In the CMS system we calculate the angle
$\theta^*_S$ between the sphericity axis of the $B$ candidate and that
of the remaining particles in the rest of the event (ROE).  While
$|\cos\theta^*_S|$ is highly peaked near 1 for continuum background,
it is nearly uniformly distributed for $B\bar{B}$ events. We require
$|\cos\theta^*_S|<0.9$, eliminating $58\%$ of the continuum events.
We also employ an event-shape Fisher discriminant in the
maximum-likelihood fit (described below) from which we extract the CPV
measurements.  This variable is defined as ${\cal F}=0.53 - 0.60 L_0 +
1.27 L_2$, where $L_j\equiv\sum_{i \in {\rm ROE}} |{\bf p}^*_i| |\cos
\theta^*_i|^j$, $\bf p^*_i$ is the momentum of particle $i$ in the
CMS system and $\theta^*_i$ is the angle between ${\bf
p}^*_i$ and the sphericity axis of the $B$ candidate.

\begin{figure}[!tbp]
\begin{center}
\includegraphics[width=0.49\linewidth]{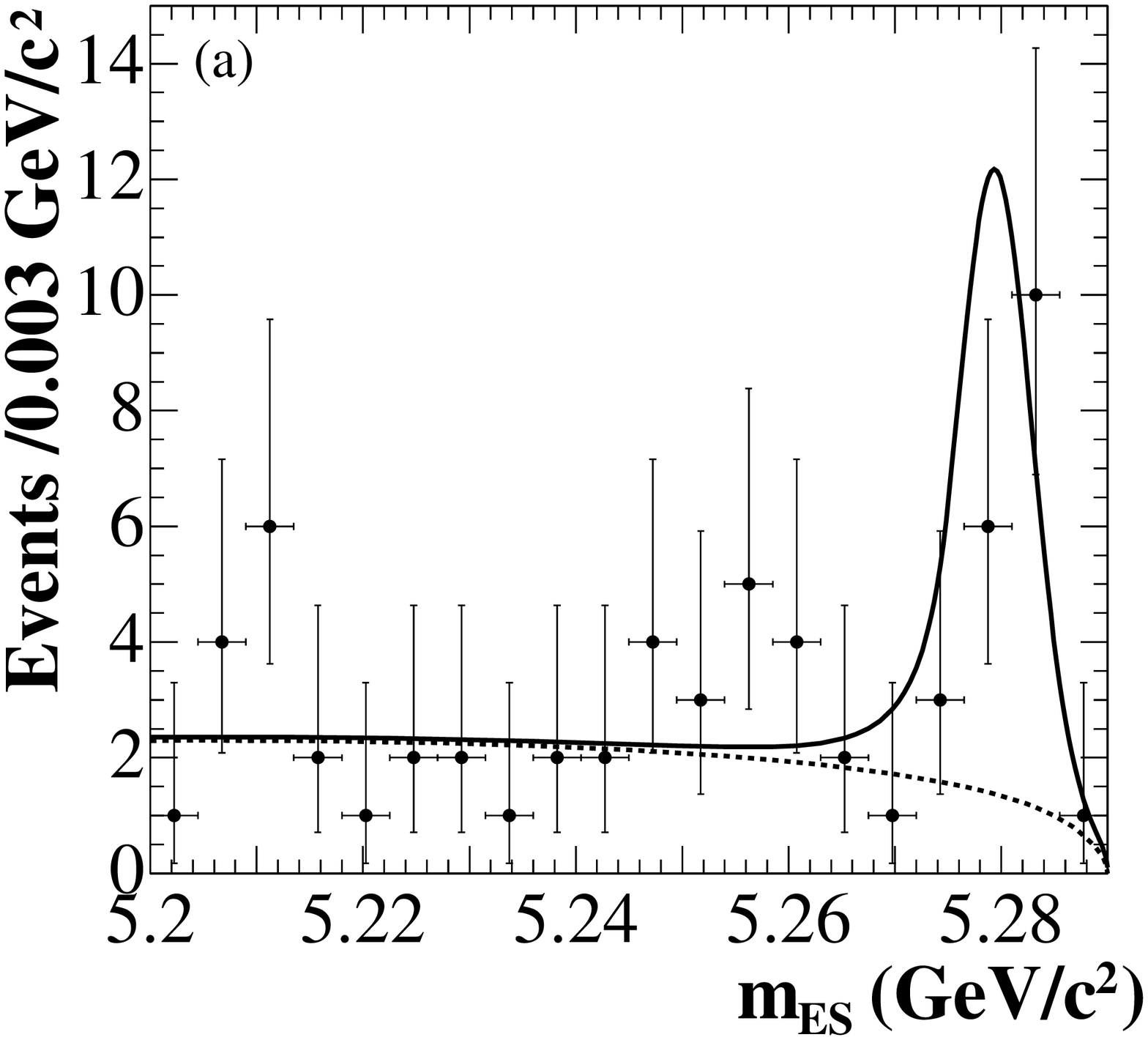}
\includegraphics[width=0.49\linewidth]{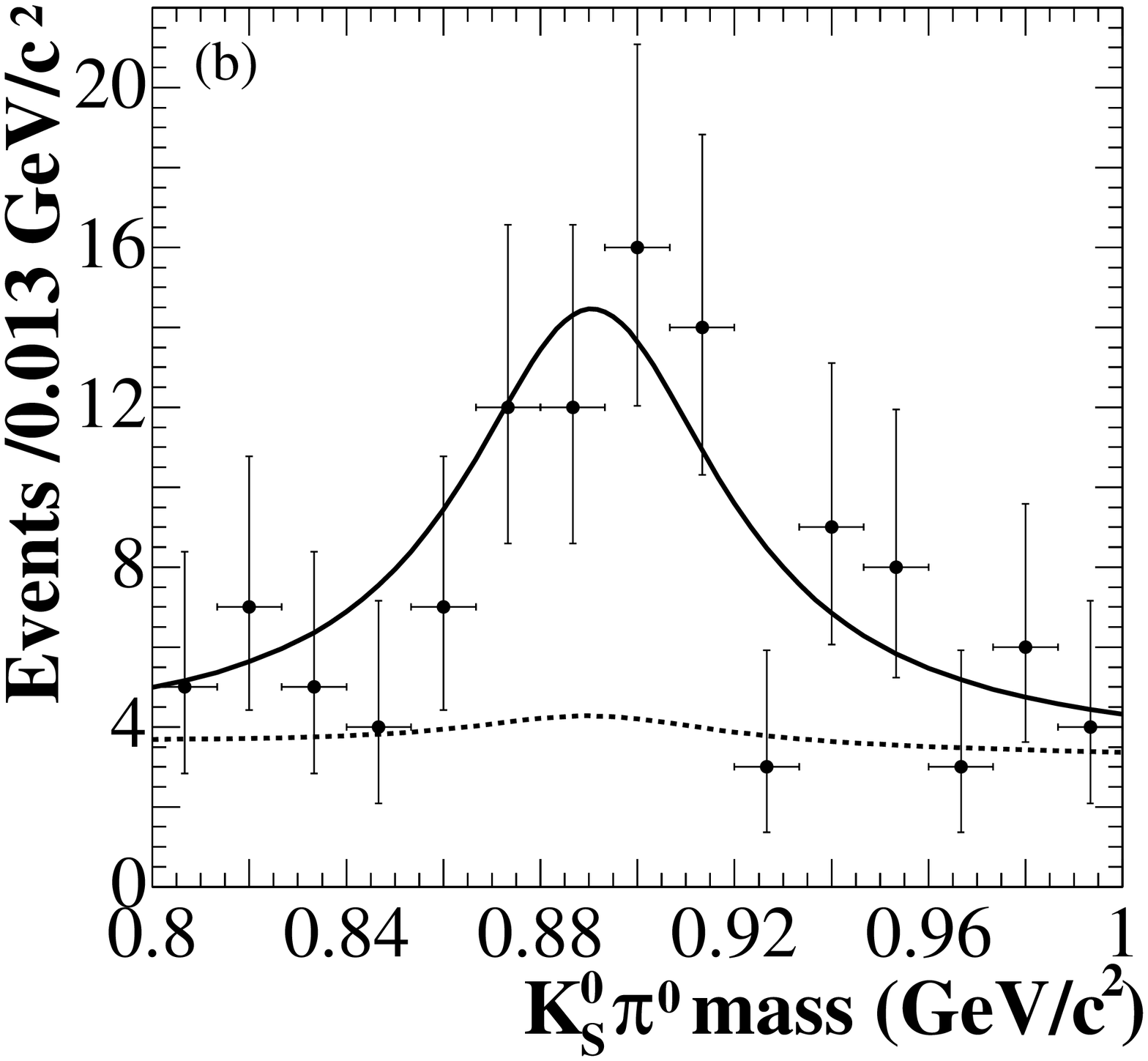}
\caption{Distribution of (a) $\mes$ and (b) $M_{\Kstar}$ for events
  enhanced in signal decays.  The dashed and solid curves represent
  the background and signal-plus-background contributions,
  respectively, as obtained from the maximum likelihood fit to the
full data sample. The
  selection technique is described in the text.}
\label{fig:prplots}
\end{center}
\end{figure}

The above selections yield 1916 $\TheDecay$ candidates.  We extract
our measurements from this sample using an unbinned maximum-likelihood
fit to kinematic (\mes, \DeltaE, and \Kstar\ mass), event shape
(\fish), flavor tag, and time structure variables (described below).
As input to the fit, we parameterize the probability distribution
functions (PDF) describing the observables of signal and $\BB$
background events using either more copious fully-reconstructed $B$
decays in data or simulated samples.  For the continuum background, we
select the functional form of the PDFs describing each fit variable in
data using the sideband regions of the other observables where the
$q\bar{q}$ background dominates.  We include these regions in the
fitted sample and simultaneously extract the parameters of the
background PDFs along with the CPV measurements. We fit $105\pm14$
signal and $19\pm15$ other $B$ decays in the selected sample. This
signal yield is consistent with expectations from the previous
measurements of the branching
fractions\cite{CLEOBRPRL,BabarBRPRL,BelleBR}. Figure~\ref{fig:prplots}
displays the \mes{} and $M_{\Kstar}$ distributions for signal-enhanced
sub-samples of these events, selected using the PDFs employed in the
fit (see below).

For each \Bztokstargamma\ candidate, we examine the remaining tracks
and neutral particles in the event to determine if the other $B$ in
the event, \Btag, decayed as a \Bz\ or a \Bzb\ (flavor tag).
Time-dependent CPV asymmetries are determined by
reconstructing the distribution of the proper decay time difference,
$\deltat\equiv t_{\CP}-t_{\rm tag}$.  At the $\Upsilon(4S)$ resonance,
the distribution of \deltat\ follows
\begin{eqnarray}
P^{\Bz}_{\Bzb}(\deltat)= &&\frac{e^{-|\deltat|/\tau}}{4\tau}[1\pm \label{eqn:td}\\
&&(S_f \sin{(\deltat \deltamd)}-C_f\cos{(\deltat \deltamd)})], \nonumber
\end{eqnarray}
where the upper (lower) sign corresponds to \Btag\ decaying as \Bz\
(\Bzb), $\tau$ is the \Bz\ lifetime, \deltamd\ is the mixing
frequency, and $S_f$ and $C_f$ are the magnitude of the mixing-induced
and direct CPV asymmetries, respectively. As stated above, in the SM
we expect $\skstargamma\approx 2 (m_s/m_b) \sin 2\beta \approx 0.05$.
We expect $\ckstargamma=-A_{\Kstar\gamma}$, the direct \CP\
asymmetry measured in the self-tagging and more copious
$\Bztokstargamma (\Kstar \ra \Kp\pim)$ decay.

We use a neural network to determine the flavor, $T$, of the $B_{\rm
tag}$ meson from kinematic and particle identification
information\cite{ref:Sin2betaPRD}. Each event is assigned to one of
five mutually exclusive tagging categories, designed to combine flavor
tags with similar performance and $\deltat$ resolution.  We
parameterize the performance of this algorithm in a data sample
($B_{\rm flav}$) of fully reconstructed $\Bz\to D^{(*)-}
\pip/\rho^+/a_1^+$ decays. The average effective tagging efficiency
obtained from this sample is
$Q=\Sigma_c\epsilon^c_S(1-2w^c)^2=0.288\pm 0.005$, where
$\epsilon^c_S$ and $w^c$ are the efficiency and mistag probabilities,
respectively, for events tagged in category $c$.  In each tagging
category, we extract the fraction of events ($\epsilon^c_{q\bar{q}}$)
and the asymmetry in the rate of $\Bz$ and $\Bzb$ tags in the
continuum background events in the fit to the data.

We compute the proper time difference \deltat{} from the known boost
of the \epem{} system and the measured $\deltaz=\zrec-\ztag$, the
difference between the reconstructed decay vertex positions of the
\Bztokstargamma{} and \Btag{} candidate along the boost direction
($z$).  A description of the inclusive reconstruction of the \Btag{}
vertex using tracks in ROE is given in
\cite{ref:Sin2betaPRD}. Replicating the vertexing technique developed
for \Bztokspiz\ decays\cite{BaBarKsPi0}, we determine the decay point
\zrec\ for \TheDecay\ candidates from the intersection of the \KS{}
trajectory with the interaction region.  This is accomplished by
constraining the \B{} vertex to the interaction point (IP) in the
plane transverse to the beam, which is determined in each run
from the spatial distribution of vertices from two-track events.  We
combine the uncertainty in the IP position, which follows from the
size of the interaction region (about \unit[200]{$\mu$m} horizontal
and \unit[4]{$\mu$m} vertical), with the RMS of the transverse \B{}
flight length distribution (about \unit[30]{\microns}) to assign an
uncertainty to the IP constraint.

Simulation studies indicate that \TheDecay\ decays exhibit properties
which are characteristic of the IP vertexing technique, namely that
the per-event estimate of the error on $\deltat$, $\sigma_{\deltat}$,
reflects the expected dependence of the \zrec{} resolution on the
$\KS$ flight direction and the number of SVT layers traversed by its
decay daughters.  Though the fit extracts \ckstargamma\ from all
flavor tagged signal decays, we only allow $68\%$ of these events
contribute to the measurement of \skstargamma. This subset consists of
candidates which are composed of \KS\ decays with at least one hit in
the SVT on both tracks and pass the quality requirements of
\unit[$\sigma_{\deltat}<2.5$]{ps} and \unit[$|\deltat|<20$]{ps}.  For
$66\%$ of this subset, both tracks have hits in the inner three SVT
layers, which results in a mean \deltat\ resolution that is comparable
to decays with the vertex directly reconstructed from charged
particles originating at the $B$ decay
point~\cite{ref:Sin2betaPRD}. In the remainder of the subset, the
resolution is nearly two times worse.

We obtain the PDF for the time-dependence of signal decays from the
convolution of Eq.~\ref{eqn:td} with a resolution function ${\cal
R}(\delta t \equiv \deltat -\deltat_{\rm true},\sigma_{\deltat})$.
The resolution function is parameterized as the sum of a `core' and a
`tail' Gaussian function, each with a width and mean proportional to
the reconstructed $\sigma_{\deltat}$, and a third Gaussian centered at
zero with a fixed width of
\unit[$8$]{ps}~\cite{ref:Sin2betaPRD}. Using simulated data, we have
verified that the parameters of ${\cal R}(\delta t, \sigma_{\deltat})$
for \Bztokstargamma\ decays and the $\BB$ backgrounds are similar to
those obtained from the $B_{\rm flav}$ sample, even though the
distributions of $\sigma_{\deltat}$ differ considerably. Therefore, we
extract these parameters from a fit to the $B_{\rm flav}$ sample. We
find that the \deltat{} distribution of continuum background
candidates is well described by a delta function convoluted with a
resolution function with the same functional form as used for signal
events. We determine the parameters of the background function in the
fit to the \TheDecay\ dataset.

To extract the CPV asymmetries we maximize the logarithm of the
likelihood function
 {\small\begin{eqnarray*}
  {\cal L}(\sf,\cf,N_h,f_h,\epsilon_{q\bar{q}}^c, \vec{\alpha})= &
 \frac{e^{-(N_S+N_{\BB}+N_{q\bar{q}})}}{(N_S+N_{\BB}+N_{q\bar{q}})\,!}  \times        \mbox{}\nonumber\\
 & \prod_{i \in \mathrm{w/\,}  \deltat} 
   [ N_S f_S \epsilon^{c}_S{\cal P}_S(\vec{x}_i,\vec{y}_i;\sf,\cf) +   \mbox{} \nonumber\\
 & N_{\BB} f_{\BB} \epsilon^{c}_{\BB}{\cal P}_{\BB}(\vec{x}_i,\vec{y}_i) +  \mbox{}\nonumber\\
 & N_{q\bar{q}} f_{q\bar{q}} \epsilon^{c}_{q\bar{q}} {\cal P}_{q\bar{q}}(\vec{x}_i,\vec{y}_i;\vec{\alpha}) 
 ] \times   \mbox{}  \nonumber\\ 
 &  \prod_{i \in  \mathrm{w/o\,} \deltat}
   [ N_S (1-f_S) \epsilon^{c}_S {\cal P}'_S(\vec{y}_i;\cf) +   \mbox{}\nonumber\\
 & N_{\BB} (1-f_{\BB}) \epsilon^{c}_{\BB} {\cal P}'_{\BB}(\vec{y}_i) +   \mbox{}\nonumber\\
 &    N_{q\bar{q}} (1-f_{q\bar{q}}) \epsilon_{q\bar{q}}^{c} {\cal P}'_{q\bar{q}}(\vec{y}_i;\vec{\alpha}) ],  \mbox{} \nonumber
 \end{eqnarray*}}
\noindent{}where the second (third) factor on the right-hand side is
the contribution from events with (without) $\deltat$ information.
The vectors $\vec{x}_i$ and $\vec{y}_i$ represent the time-structure
and remaining observables, respectively, for event $i$. The
PDFs ${\cal
P}_h(\vec{x}_i,\vec{y}_i)=P_h(\mes_i)P_h(\DeltaE_i)P_h({\cal
F}_i)P_h(M_{\Kstar,i}) P_h^{c_i}(\deltat_i|\sigma_{\deltat,i},T_i)$ and
${\cal P'}_h(\vec{y}_i)=P_h(\mes_i)P_h(\DeltaE_i)P_h({\cal
F}_i)P_h(M_{\Kstar,i}) P_h^{c_i}(T_i)$ are the products of the PDFs
described above for hypothesis $h$ of signal ($S$), $\BB$ background
($\BB$), and continuum background ($q\bar{q}$). Along with the CPV
asymmetries \sf\ and \cf, the fit extracts the yields $N_S$,
$N_{\BB}$, and $N_{q\bar{q}}$, the fractions of events with $\deltat$
information $f_S$ and $f_{q\bar{q}}$, and the parameters
$\vec{\alpha}$ which describe the background PDFs. We determine
$\epsilon^c_B$ and $f_{\BB}$ in simulated generic \BB\ decays.

The fit to the data sample yields \mbox{$\skstargamma = 0.25 \pm 0.63
\pm 0.14$} and \mbox{$\ckstargamma = -0.57 \pm 0.32 \pm 0.09$}, where
the uncertainties are statistical and systematic, respectively.  The
fit reports a correlation of $1\%$ between these parameters. The
result for $\ckstargamma$ is consistent with a fit that does not
employ $\deltat$ information.  Since the present measurements of
$A_{\Kstar\gamma}$\cite{BabarBRPRL,BelleBR} are consistent with zero,
we also fit the data sample with $\ckstargamma$ fixed to zero and
obtain $\skstargamma=0.25\pm 0.65\pm 0.14$.

\begin{figure}[!tbp]
\begin{center}
\includegraphics[width=0.9\linewidth]{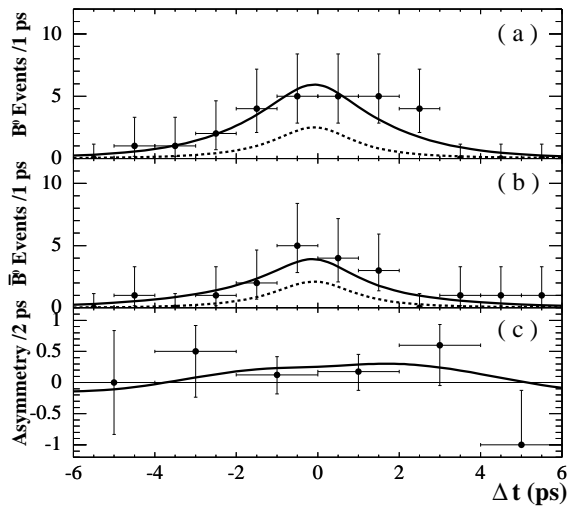}
\end{center}
\caption{ Distributions of $\deltat$ for events enhanced in signal
decays with $B_{\rm tag}$ tagged as (a) $\Bz$ or (b) $\Bzb$, and (c)
the resulting asymmetry ${\cal A}_{\Kstar\gamma}(\deltat)$.  The
dashed and solid curves represent the fitted background and
signal-plus-background contributions, respectively, as obtained from
the maximum likelihood fit. The raw asymmetry projection corresponds
to approximately $38$ signal and $19$ background events.}
\label{fig:dtplot}
\end{figure}

The event selection criteria employed to isolate signal-enhanced
samples displayed in Figure~\ref{fig:prplots} are based on a cut on
the likelihood ratio $R={\cal P}_S/({\cal P}_{S}+{\cal P}_{BB}+{\cal
P}_{q\bar{q}})$ calculated without the displayed observable. The
dashed and solid curves indicate background and signal-plus-background
contributions, respectively, as obtained from the fit, but corrected
for the selection efficiency of $R$.  Figure~\ref{fig:dtplot} shows
distributions of $\deltat$ for $\Bz$- and $\Bzb$-tagged events, and
the asymmetry ${\cal A}_{\Kstar\gamma}(\deltat) = \left[N_{\Bz} -
N_{\Bzb}\right]/\left[N_{\Bz} + N_{\Bzb}\right]$ as a function of
$\deltat$, also for a signal-enhanced sample.

We consider several sources of systematic uncertainties related to the
level and possible asymmetry of the background contribution from
generic $B\bar {B}$ decays. We estimate the impact of potential biases
in the determination of the \BB\ background rate to lead to a
systematic uncertainty of $0.04$ ($0.05$) on \skstargamma\
(\ckstargamma). We estimate an uncertainty of $0.12$ ($0.03$) due to
potential CPV asymmetries in the $\BB$ backgrounds and $0.02$ ($0.06$)
due to possible asymmetries in the rate of \Bz{} versus \Bzb{} tags in
continuum backgrounds. We quantify possible systematic effects due to
the vertexing method in the same manner as Ref.\cite{BaBarKsPi0},
estimating systematic uncertainties of $0.04$ ($0.02$) due to the
choice of resolution function, $0.04$ ($<0.01$) due to the vertexing
technique, and $0.03$ ($0.01$) due to possible misalignments of the
SVT.  Finally, we include a systematic uncertainty of $0.02$ ($0.02$)
due to tagging asymmetries in the signal and $0.02$ ($0.02$) due to
imperfect knowledge of the PDFs used in the fit.

In summary, we have performed a measurement of the time-dependent CPV
asymmetry \skstargamma\ and the direct-\CP\ violating asymmetry
\ckstargamma\ from \TheDecay\ decays. Our measurement is consistent
with the SM expectation of very small CPV asymmetries.

We are grateful for the excellent luminosity and machine conditions
provided by our \pep2\ colleagues, 
and for the substantial dedicated effort from
the computing organizations that support \babar.
The collaborating institutions wish to thank 
SLAC for its support and kind hospitality. 
This work is supported by
DOE
and NSF (USA),
NSERC (Canada),
IHEP (China),
CEA and
CNRS-IN2P3
(France),
BMBF and DFG
(Germany),
INFN (Italy),
FOM (The Netherlands),
NFR (Norway),
MIST (Russia), and
PPARC (United Kingdom). 
Individuals have received support from CONACyT (Mexico), A.~P.~Sloan Foundation, 
Research Corporation,
and Alexander von Humboldt Foundation.


\begin{thebibliography}{99}
\bibitem{ref:cc}
Unless explicitly stated, charge conjugate decay modes are assumed throughout this paper.
\end{thebibliography}
\end{document}